\PassOptionsToPackage{table}{xcolor}
\documentclass[manuscript]{acmart}
\usepackage{graphicx} 
\usepackage{booktabs}
\usepackage{caption}
\usepackage{subcaption}
\usepackage{framed}
\usepackage{listings}  
\usepackage{array}
\usepackage{todonotes}
\usepackage{comment}
\usepackage{multirow}
\usepackage{xcolor}
\usepackage{xspace}
\usepackage{forest}

\tikzset{
        my node/.style={
            draw=gray,
            inner color=gray!5,
            outer color=gray!10,
            thick,
            minimum width=1cm,
            rounded corners=3,
            text height=1.5ex,
            text depth=0ex,
            font=\sffamily
        }
    }

\definecolor{Gray}{gray}{0.95}
\definecolor{Gray2}{gray}{0.92}

\newcommand{\numVenues}{50\xspace} 

\newcommand{\papersRelevant}{390\xspace} 
\newcommand{\papersExcluded}{132\xspace} 
\newcommand{\papersExtract}{79\xspace} 

\title[The Green Software Landscape: A Systematic Mapping Study]{The Green Software Landscape: A Systematic Mapping Study on Evolution, Applications, Software Lifecycle, and Best Practices}

\author{Max Hort}
\affiliation{
  \institution{Simula Research Laboratory}            
  \country{Norway}                   
}
\email{maxh@simula.no}          
\author{Maria Kechagia}
\affiliation{
  \institution{University of Athens}           
  \country{Greece}                   
}
\email{makechag@ba.uoa.gr}         
\author{Federica Sarro}
\affiliation{
  \institution{University College London}           
  \country{United Kingdom}                  
}
\email{f.sarro@ucl.ac.uk}        

\begin{document}

\begin{abstract}
Energy consumption and climate change have made sustainability critical in Software Engineering (SE), driving the emergence of Green SE. Over the past 15 years, numerous solutions for sustainable software systems have been published by the SE community, offering a rich resource for analyzing the field's evolution.
		
		To explore this, we conducted a systematic mapping study of Green SE research published between 2010 and 2024. 
		We collected \papersRelevant  publications, categorizing them by application domain (e.g., mobile, cloud, AI) and research type (e.g., optimisation study, benchmarking, literature review Additionally, we analyzed a representative subset of \papersExtract papers to classify the key elements—such as hardware, measurement, stability, and replicability—considered during energy measurement experiments.
		
		
		Our findings indicate that SE conferences host the majority of energy-related literature. Notably, Green SE studies surged in popularity starting in 2023, largely driven by AI-related publications. Optimization and benchmarking emerged as the most prevalent research types. Ultimately, we aim to inform the SE community about current approaches to energy concerns, highlight critical experimental practices, and advocate for continued action toward more sustainable software engineering.
\end{abstract}
\maketitle

\section{Introduction}
In 2010, Easterbrook~\cite{easterbrook2010climate} described climate change as a ``grand software challenge'' since software,
which itself consumes energy,
can be used to address climate change.
Specifically, software can be used to measure energy consumption and identify energy-hungry software modules,
facilitating decision making for reducing energy consumption.
The interest in Green Software Engineering (SE) has been growing fast over the last 15 years, leading to the creation of specialised workshops addressing sustainable software, including the ICSE co-located event GREENS (International Workshop on Green and Sustainable Software),\footnote{https://greensworkshop.github.io/} which was launched in 2012.
There is a great deal of publications regarding Green SE
in all top SE venues, including ICSE, FSE, ASE, MSR, TSE, and TOSEM.
For instance, ICSE had 39 Green SE publications since 2010 and the GREENS workshop over 50 between 2012-2024.

Furthermore, recently, concerns regarding software sustainability have been raised in dedicated keynotes at leading SE conferences.
For instance, consider the ICSE 2025 keynote %
by Professor Lago on ``Software Sustainability and its Engineering: How far have we come?". \footnote{\url{https://conf.researchr.org/details/icse-2025/icse-2025-main-icse-plenaries/4/Patricia-Lago-Keynote-Title-Software-Sustainability-and-its-Engineering-How-far-ha}}

Additionally, in 2024, FSE presented a new collocated event entitled ``2030 Software Engineering workshop''
to discuss the future of software engineering, including green-aware software development.\footnote{https://conf.researchr.org/home/2030-se}
In particular, 6 out of 59 papers were relevant to software sustainability, pointing out the importance of green software.

Although the number of Green SE publications within the SE community is steadily increasing, the field remains fragmented. Research outputs are distributed across multiple venues, including journals, workshops, and conferences, each with different goals and application targets~\cite{hindle2016green}.

These publications address multiple facets of sustainability, including social, environmental, technical, and economic dimensions~\cite{becker2015sustainability,lago2015framing,mcguire2023sustainability}.
Here, we focus specifically on the environmental dimension,
with a particular emphasis on energy-related concerns,
namely the energy consumed by software systems and applications,
resulting in the emission of CO2
and having an impact on global warming~\cite{mourao2018green,amsel2011toward,cook2016consensus}.
Overall, the aim of this study is to inform the SE community about
the ways in which energy-related concerns are currently addressed within the SE domain,
and call for further actions towards Green SE.

Specifically, we aim to identify the leading software engineering venues that address energy-related concerns
to understand how much the SE community is green-aware.
Furthermore, we wish to distil trends across these venues, for example, whether certain topics are primarily discussed in workshops rather than in the main tracks of SE conferences.
Additionally, our goal is to identify
for {\it which} software applications researchers measure energy consumption and {\it how},
as well as {\it what} key elements researchers
consider in their experiments.
To this end, we perform a systematic mapping study of Green SE work published in the most popular and renowned SE journals and conferences, and their co-located events.

There exist several prior mapping studies on sustainable software engineering \cite{penzenstadler2014systematic,marimuthu2017software,lee2024survey}.
However, their scope does not go beyond 2021, with 101 collected publications at most. 
Moreover, the search process considered did not focus on the venues  themselves, but rather started with search databases.
To the best of our knowledge,
we are the first to consider the most comprehensive set of publications, i.e., \papersRelevant publications over 15 years (2010-2024)
to provide classifications of the publications examined from the following three aspects:
applications, methods, and experiments.

To this end, we first examine all top SE venues in the field (RQ1).
Then, we categorise the publications per field of applications considered
e.g., mobile, artificial intelligence, cloud, and so on (RQ2).
Additionally, we discuss the types of studies (e.g., optimisation, benchmarking) of the publications examined (RQ3).
Finally, we summarise the key elements
considered in the design protocols
of the empirical studies of the publications examined (RQ4).

The main findings of our work are as follows:
\begin{itemize}
    \item Most publications on energy-related concerns
    appear in top SE conferences (183 out of \papersRelevant, 47\%)  and workshops (34\%), whereas only a smaller number of publications appear
    in the proceedings of journals (19\%).  
    \item Mobile computing is the application domain in which energy-related concerns have been most frequently investigated in the SE literature over the past 15 years.
    Additionally, Artificial Intelligence (AI) represents
    a fast-growing application domain
    where researchers investigate energy-related issues.
    \item Optimisation and benchmarking studies
    are the most prevalent types of research in the field of Green SE.
    \item The most critical elements that researchers consider
    when conducting experiments on measuring energy include: hardware requirements, energy-measurement tools, experiments' stability and replication, as well as benchmarks.
\end{itemize}

The remainder of the paper is structured as follows.
The background is outlined in Section~\ref{section:background}.
Our study's design is outlined in Section~\ref{section:method}.
Section~\ref{section:results} presents the results of our investigation, and Section~\ref{section:implications} provides further insights and discussion on the implications of our findings on the field of Green SE.
Section~\ref{section:threats} discusses the threats to validity of our study.
Related work is discussed in Section~\ref{section:rw}.
Section~\ref{section:conclusion} concludes our study.

\section{Background}
\label{section:background}
In this section, we present the key context,
terms, and definitions
used in our survey.

\subsection{Green Software Engineering (SE)}
\label{greense}
Green SE has emerged as a research area focused on
reducing energy consumption across software engineering tasks and processes.
Green SE is closely related to the concept of {\it software sustainability}.
Becker et al.~\cite{becker2015sustainability}
identified several definitions of the term,
with the most prevalent referring to ``the capacity to endure''.
According to Penzenstadler et al.~\cite{penzenstadler2014systematic},
sustainability is often described in
the software-engineering literature as a non-functional requirement or a software quality attribute.
As revealed by the survey of Lee et al.~\cite{lee2024survey},
most researchers define sustainable software as
software that minimizes negative impacts on society, the economy, and the environment
throughout the software development life cycle~\cite{DNK10,CMB14}.
This implies that software engineering should continuously
seek to reduce resource consumption,
including energy usage,
in order to mitigate environmental impact and promote social and economic
sustainability~\cite{CP15}.
We refer to this definition of sustainability throughout this survey when we use the terms Green software or Green SE.

\subsection{Systematic Mapping Studies}

Systematic mapping studies aim to
give an overview of a research area
through classification and
the frequency of contributions
in relation to the categories identified for that classification~\cite{KC07,PFM08,PVK15}.
According to Petersen et al.~\cite{PFM08},
a software engineering systematic map refers to a structured method that
builds a classification,
and organises, based on that classification,
a software engineering field of interest, i.e., Green SE in our case.
The analysis focuses on
the frequencies of the publications under examination
that belong to the categories identified within a classification.
Here, we opted for a systematic mapping study
since we wish to investigate
the evolution of the Green SE research field over the last 15 years,
including studies from the recently emerging subfield of Green AI.

Consider that systematic mapping studies differ from
systematic literature reviews (SLRs).
According to Kitchenham et al.~\cite{KBB10},
there are differences
between the two types of studies
with respect to the research questions, search
process, search strategy requirements, quality evaluation, and
results.
Particularly, the research questions in mapping studies are typically broad
since the research topics aim to reveal research trends
(e.g., the evolution of publications,
the topics covered in the literature).
By contrast, systematic reviews are designed to aggregate evidence,
and answer specific questions.
This means that systematic reviews serve very specific goals
formulated (e.g., the assessment of the effectiveness of a specific technique).

\subsection{Software Engineering Body of Knowledge (SWEBOK)}

According to Lee et al.~\cite{lee2024survey},
sustainability should be considered throughout the entire software
development life cycle to enable continuous ``green software'' practices.
This perspective is consistent with the principles of the
Software Engineering Body of Knowledge (SWEBOK),\footnote{We consider
the latest version of SWEBOK (Version 4.0), available at:
https://www.computer.org/education/bodies-of-knowledge/software-engineering.}
an internationally recognized guide to the software engineering
discipline~\cite{BDA99}.
SWEBOK serves as a reference for both research and education,
covering a broad set of knowledge areas that constitute the foundation
of software engineering.
The current version comprises 18 knowledge areas,\footnote{A complete
list of SWEBOK knowledge areas is available at:
https://www.computer.org/education/bodies-of-knowledge/software-engineering/topics.}
which together provide a comprehensive framework for understanding and
supporting the various aspects of software development and management.
Several previous surveys have adopted SWEBOK to organize and summarize
the Green Software Engineering literature~\cite{mourao2018green,
penzenstadler2014systematic,marimuthu2017software,
wolfram2017sustainability,lee2024survey}.

In this study, we also adopt SWEBOK as the conceptual framework for
classifying the selected studies.
Specifically, we group the literature into four broad knowledge areas:
\emph{Software Requirements}, \emph{Software Architecture},
\emph{Software Construction}, and \emph{Software Testing}.
Together, these areas cover the core phases of the software development
life cycle, while also encompassing cross-cutting concerns related to
software quality, processes, methods, and management.
This classification enables a comprehensive examination of how
energy-related concerns are addressed throughout software engineering
practice.
In the following paragraphs, we briefly describe each category.

\paragraph{Software Requirements.}
This knowledge area concerns the elicitation, analysis, specification,
validation, and management of software requirements.
In the context of Green Software Engineering, it includes studies that
investigate how energy-related concerns are identified and formalized
as software requirements, and how such requirements are managed and
evolved throughout the software development life cycle.

\paragraph{Software Architecture.}
This knowledge area addresses the high-level organization of a software
system, including its components, their externally visible properties,
and the relationships among them.
Studies in this category examine architectural decisions, design
patterns, and architectural strategies that influence the energy
efficiency of software systems.

\paragraph{Software Construction.}
This knowledge area focuses on the development of software through
coding, verification, and related implementation activities.
Specifically, we consider programming practices, coding guidelines,
verification techniques, and development strategies.
Within the context of Green Software Engineering, this category
encompasses studies that investigate how implementation choices,
programming languages, frameworks, and development environments affect
software energy consumption.
We also include studies on optimization techniques for reducing energy
consumption.
Throughout this survey, we use the terms \emph{software construction}
and \emph{software development} interchangeably.

\paragraph{Software Testing.}
This knowledge area covers the systematic evaluation of software to
assess its quality and correctness through activities such as unit,
integration, and system testing.
In this category, we include studies that propose methods and tools for
measuring and analyzing energy consumption during testing activities.

\section{Study Design}
\label{section:method}
We conduct a
{\it systematic mapping study}
providing an overview of
the Green SE research area
through classification and metrics regarding the publications identified~\cite{KC07,PFM08,PVK15}. 
This section outlines the protocol we used for this systematic mapping study.
 This protocol refers to a predefined plan
that defines the research questions
and the procedures followed to achieve the goals of our study~\cite{PFM08,PVK15}.

\subsection{Research Questions}
The purpose of our study is to
inform the SE community
about the ways in which energy-related concerns
are currently addressed within the SE domain,
and call for further actions towards Green SE.
To do so, we analyse data
extracted from an extensive collection of
papers published
in leading SE venues.
We formulate the following four research questions:

\begin{itemize}
    \item {\bf RQ1: 
    To what extent have the main SE  research venues addressed energy-related concerns from 2010 to 2024?}\\
    {\it Motivation.}
    We aim to answer RQ1
    to understand and highlight
    the ``attitude'' of the SE community
    towards emerging green software aspects.
    This is particularly important given that climate change demands action across all sectors of modern life, and software engineering represents a key domain for providing practical solutions~\cite{easterbrook2010climate}.\\
    {\it Goal.} Addressing RQ1 will provide a snapshot of the current state in the field of Green SE and encourage additional research endeavours
    focused on environmentally conscious software development.\\
    {\it Methods.} To answer RQ1, two authors cross-check the publications of all top SE venues (see Table~\ref{table:venues}) from 2010 to 2024
    that refer to sustainability.
    Section~\ref{selection_process} presents the methodology we used to answer RQ1.
    We are interested in the venue type and the amount of relevant publications.
    We have been inspired for that by the systematic mapping study of Ferreira et al.~\cite{ferreira2021software}.
    Furthermore, similar to other existing surveys~\cite{marimuthu2017software,mourao2018green,penzenstadler2014systematic,verdecchia2023systematic},
    we extract data and categorise the publications collected.\\ 
    \item {\bf RQ2: Which application domains are the most investigated with respect to energy-related concerns?} \\ {\it Motivation.} We aim to answer RQ2 to examine {\it which} application domains mostly acknowledge the importance of energy-aware software.
    Such an investigation is critical, as modern software applications,
    particularly AI-based ones,
    consume substantial amounts of energy~\cite{strubell2020energy},
 contributing to significant environmental costs.
    Understanding which application domains of SE already address energy consumption will allow us to identify critical SE domains that require further efforts towards Green SE. For example, we aim to unveil whether application domains that require large amounts of energy actually consider any optimisation approaches to reduce energy consumption.\\
    {\it Goal.} Answering RQ2 will summarise which fields of SE applications are the most green-aware. Consequently, RQ2 will highlight the fields that require more efforts towards green-aware development.\\
    {\it Methods.} To answer RQ2, two authors categorise the publications collected based on the applications' field, such as assessing energy consumption in mobile applications.
    For identifying the application fields,
    we consider the official ACM Computing Classification System.\footnote{\url{https://dl.acm.org/ccs}}
    Section~\ref{sec-analysis} details our data analysis.\\
    \item {\bf RQ3: What types of research approaches are exploited to tackle energy-related concerns in different SE tasks?}\\
    {\it Motivation.} We wish to address RQ3 to understand {\it how} different fields of applications address energy-related concerns in SE.
    Namely, we investigate the different types of research and tasks considered
    for promoting energy-aware software development in different fields of SE applications.\\
    {\it Goal.} Answering RQ3 will provide a summary of existing approaches used in SE research for reducing energy consumption for specific SE tasks.\\
    {\it Methods.} To answer RQ3, two authors manually examine each publication to gather its research type (e.g., empirical study) and task (e.g., optimisation).
    To categorise the tasks, we have considered the official classification provided by IEEE's SEWBOK.\footnote{\url{https://www.computer.org/education/bodies-of-knowledge/software-engineering/}}
    Section~\ref{sec-analysis} presents our methodology.\\
     \item {\bf RQ4: What are the key elements of the experimental design protocols considered in SE studies
     that address energy-related concerns?}\\
    {\it Motivation.}
    Empirical studies in Green SE,
    should rely on well-defined experimental design protocols
    to ensure the accuracy and reproducibility of the results.
    Key elements including hardware configuration, execution environment stability, repetition strategies, and energy measurement approaches are critical for shaping reliable experimental outcomes.
    As Green SE research increasingly involves complex and computationally intensive techniques,
    including AI-based software systems,
    a systematic understanding of
    how experiments are designed and conducted
    is essential for advancing the Green SE field.
    
    {\it Goal.}
    The goal of RQ4 is to identify and summarise
    the key elements of the experimental design protocols considered
    in empirical Green SE studies.
    Analysing how experiments are designed and executed will provide us with
    a structured overview of common design practices and methodological choices,
    and assist us to make suggestions to the SE community for
    a more robust and reproducible future research in Green SE.
    
    {\it Methods.}
    To address RQ4, we analyse the experimental design protocols described in a representative set of empirical studies included in this review, comprising a total of \papersExtract publications.
    Key design elements were extracted through a systematic examination of experimental setups, procedures, and methodological descriptions.
    Based on our analysis, we classify the elements identified into five main categories: 1. Hardware, 2. Stability, 3. Repetitions, 4. Measurement, and 5. Datasets.
  
    Section~\ref{filter-rq4} outlines the filtering conducted for RQ4 and Section~\ref{sec-analysis} explains the process for the classification of the elements.
\end{itemize}

\subsection{Search Strategy}
\label{section:search_procedure}

In the following,
we outline the criteria used to select the venues
included in our study
and describe the search procedure
to identify relevant publications.

\label{section:protocol}
\begin{table}[t]
\caption{Venues considered in our search procedure.}
\label{table:venues}
\centering
\rowcolors{1}{}{Gray2}
\begin{tabular}{ll}
\toprule
Acronym & Name \\ \midrule
ICSE & International Conference on Software Engineering \\
ESEC/FSE & International Conference on the Foundations of Software Engineering \\
ASE & International Conference on Automated Software Engineering \\
ICST & International Conference on Software Testing, Verification and Validation \\
SANER & International Conference on Software Analysis, Evolution and Re-engineering \\
ICSME & International Conference on Software Maintenance and Evolution \\
ISSTA & International Symposium on Software Testing and Analysis \\
ESEM & International Symposium on Empirical Software Engineering and Measurement \\ 
EASE & International Conference on Evaluation and Assessment in Software Engineering \\
ISSRE &  International Symposium on Software Reliability Engineering \\
RE &  International Requirements Engineering Conference  \\

ICSA (2024-2017)  & International Conference on Software Architecture \\
WICSA (2016-2010) & Working IEEE/IFIP Conference on Software Architecture \\
\midrule
JSS & Journal of Systems and Software \\
TSE & Transactions on Software Engineering \\
TOSEM & Transactions on Software Engineering and Methodology \\
EMSE &  Empirical Software Engineering \\ 
\bottomrule
\end{tabular}
\end{table}

\begin{table}[t]
\caption{Co-Located events with Green SE publications.}
\label{table:co-located}
\centering
\rowcolors{1}{}{Gray2}
\begin{tabular}{lp{8cm}}
\toprule
Conference & Co-located Events \\ \midrule
ICSE &
AST, CHASE, E2SC, GREENS, ICPC, MiSE, MOBILESoft, PLEASE, MSR, RoSE, TechDebt, WAPI, WETSEB, WETSoM, WODA+PERTEA, SAM, SE4Science, SEAMS, SEsCPS, BoKSS, GI, CAIN, SESoS, GAS, RAIE, LLM4Code \\ 
ESEC/FSE &
DeMobile, SSBSE, SERF, EnSEmble, AIWare, 2030 Software Engineering, FoSER \\ 
ASE &
A-Mobile, SUSTAIN-SE \\ 
ICST &
ITEQS, NEXTA, ASQT, IWCT \\ 
ICSA & GREENS (2024) \\
SANER & IWBOSE, FOSE \\ 
ISSTA &
WODA, ECOOP, VORTEX, icooolps, DPA, SOAP \\ \midrule
ESEM &
MeGSuS \\
EASE & DevOps \\
ISSRE & IWSF \& SHIFT \\
RE & REFrame, RESET, REW \\

\bottomrule
\end{tabular}
\end{table}

{\bf Venue selection.}
We choose to conduct our search {\it manually} from leading venues, without relying on automated search tools such as the ACM Digital Library. This procedure is commonly adopted by other surveys in SE~\cite{vegas2023pitfalls,ferreira2021software}, which goal is to identify only {\it primary studies}
of high quality published in top-tier SE venues.
Collecting papers directly
from leading venues is a well-established practice
in SE secondary studies,
as reported by Kitchenham et al.~\cite{KBB09,KBT09}.
Although high-quality primary studies are not strictly required for conducting a systematic mapping study,
we opt to focus on top-tier venues for two main reasons:
1) To capture trends in the field of Green SE;
since high-quality publications typically represent
the most advanced research in a research area under examination~\cite{KAA21}.
2) To surface research directions and challenges
that can influence SE researchers and practitioners' communities;
since papers published in top-tier venues tend to have a greater impact across such communities~\cite{KGS23April}.

Furthermore, for completeness,
we also include in our analysis all the co-located events
of the top-tier conferences selected.
These events often reflect emerging trends and typically feature specialised workshops in the area of interest, such as Green SE.
For instance, GREENS,\footnote{\url{https://conf.researchr.org/home/icse-2026/greens-2026}} which is co-located with ICSE, offers publications highly relevant to Green SE, even though it is not  itself classified as an A/A* venue by CORE.

Following the well-established CORE ranking,\footnote{\url{https://www.core.edu.au/}}
we consider as top-tier SE venues the conferences and journals ranked as A or A*.
For each selected conference we consider its co-located events.
The conferences and journals included in our study are listed in Table~\ref{table:venues}, while Table~\ref{table:co-located} lists co-located events.

{\bf Publication retrieval.}
To identify the research studies included in our analysis,
we manually inspected the official websites of the selected top-tier conferences 
(and their co-located events), and journals.
When a paper could not be retrieved directly from a conference's website,
we consult the ACM Digital Library,\footnote{\url{https://dl.acm.org/}}
IEEE Xplore Digital Library,\footnote{\url{https://ieeexplore.ieee.org/Xplore/home.jsp}}
or DBLP,\footnote{\url{https://dblp.org/}}
where the proceedings of the conferences are typically indexed.

{\bf Search period.} Our search spans a 15-year period, from 2010 to 2024.
We believe that this time-frame can offer a comprehensive overview of 15 years of Green SE research,
including all prominent SE venues in the field.
It is also worth noting that, according to our research,
before that period,
the availability and maintenance of conference and workshop websites
is limited.
This fact lead us to set 2010
as the starting year for our search.

\subsection{Selection Criteria}
\label{section:selection_criteria}

\subsubsection{Filtering for RQ1-RQ3}
\label{filter-rq1-rq3}

In the following paragraphs, we outline the procedure employed to select the primary studies included in our analysis for RQ1-RQ3,
together with the corresponding number of studies retained at each stage of the process followed.

We set the following inclusion (IC) and exclusion criteria (EC)
to be considered for all the studies
collected within the search procedure explained in Section \ref{section:search_procedure}.
This guarantees that only relevant primary studies
will be included in our systematic mapping study.

{\bf Inclusion criteria.}
The inclusion criteria  for our systematic mapping study
are defined as follows:

\begin{itemize}
\item IC1: The study addresses topics related to software engineering. {\bf AND}
\item IC2: The study refers to sustainability (specifically, energy-related concerns).
\end{itemize}

These inclusion criteria are derived from our research questions,
since the goal of our work is to clarify
{\it how} sustainability is currently considered
within the field of software engineering.
Accordingly, the primary studies included {\it must} address
both areas considered in our survey, software engineering and sustainability (in particular, energy-related concerns).

{\bf Exclusion criteria.} The exclusion criteria for our systematic mapping study
are defined as follows:

\begin{itemize}
\item EC1: The study is an editorial, keynote, tutorial, or panel discussion.
\end{itemize}

This exclusion criterion aligns with common exclusion practices in systematic mapping studies, as indicated in related work~\cite{KAA21}.
Specifically, our aim is to include only studies
that are sufficiently complete,
present a novel contribution,
and provide an accompanying evaluation.
Studies falling under EC1
typically lack such evaluations
and are therefore excluded from our paper.

Consider that our exclusion criteria do not explicitly exclude publications written in languages other than English, as commonly done in other systematic literature reviews~\cite{KAA21,KGS23}. This is because, a priori, all studies published within top-tier SE venues should be written in English. 

Overall, we filter out \papersExcluded publications, and we keep \papersRelevant publications.

\subsubsection{Additional filtering for RQ4}
\label{filter-rq4}

For RQ4, we are only interested in a subset of publications from the \papersRelevant investigated for RQ1-RQ3.
Therefore, for RQ4, we additionally apply the following inclusion (IC) and exclusion (EC) criteria.

{\bf Inclusion criteria.} We consider two types of studies.

\begin{itemize}
\item IC3: The paper presents an ``optimisation'' study that creates methods to reduce energy consumption. {\bf OR}
\item IC4: The study is a ``benchmarking'' one. Namely, it compares different implementation choices to recommend the best one in terms of energy consumption (e.g., which collection type to use in Java).
\end{itemize}

{\bf Exclusion criteria.} The additional exclusion criteria follows.

\begin{itemize}
\item EC2: Publications with 6 or less pages. We choose this limit to ensure that the considered publications have enough space to carry out and report comprehensive empirical results and design choices. {\bf AND}
\item EC3: Hardware/cloud-based studies, as well as approaches that optimise energy consumption for testing (e.g., test suite reduction), rather than the software itself. We exclude these studies since they do not refer to or optimise the core components of software systems themselves (executed by users) for sustainability, as we are interested in here.
\end{itemize}
 
Applying the criteria mentioned leaves us with a total of \papersExtract publications from which we extract design protocols.

\subsection{Selection Process}
\label{selection_process}

To identify relevant primary studies
for inclusion in our systematic mapping study,
we begin by conducting a manual search
(see Section~\ref{section:protocol})
of the conferences and journals listed in Table~\ref{table:venues}.
In addition to these primary sources,
we identify further studies across 33 co-located events.
Table~\ref{table:co-located} presents the co-located events
that featured Green SE publications for each associated conference.\footnote{Note that from 2012 to 2023, GREENS was co-located with ICSE; however, in 2024 it was co-located with ICSA.}

The first two authors divided the set of venues under consideration (Section~\ref{section:search_procedure})
and independently applied the inclusion and exclusion criteria for RQ1-RQ3 (see Section~\ref{filter-rq1-rq3}) to identify and remove irrelevant records.
This process resulted in a collection of 521 primary studies, whose titles suggested potential relevance across all venue types, i.e., conferences, journals, and co-located events.
At this stage, the studies were selected based on the relevance of their titles to sustainability-related topics
(e.g., green, energy, power),
without requiring the presence of specific keywords.
At the end, we manually verified that the included studies {\it indeed} contained sustainability-related terms in their titles, i.e., ``green'', ``energy'', ``sustainability'', and ``power''.

After the selection of the primary studies based on their titles,
we also review their abstracts, as well as the full texts when necessary,
to assess whether the papers are still relevant to our goals, i.e., whether the papers address sustainability (in particular, energy-related concerns).
Through this selection process,
we excluded \papersExcluded studies.

Consequently, our final dataset comprises \papersRelevant primary studies, whose annual distribution is presented in Figure~\ref{fig:years}.

Finally, the first author applied the additional filtering for RQ4 (see Section~\ref{filter-rq4}),
and extracted a total of \papersExtract publications
for a further analysis.
Consider that the filtering for RQ4
was deterministic,
based on the already labelled categories
given by the two first authors for RQ2 and RQ3.
Therefore, we believe that one validation
is adequate for the process.

\subsection{Data Extraction}
For each primary study we selected,
we recorded the following information to answer RQ1-RQ3:

\begin{itemize}
    \item Source: Conference, journal, or co-located event
    \item Title: Paper title.
    \item Year: Publication year.
    \item Application domain (for RQ2): Inspired by the related work
    and the ACM's Classification.
    \item Study type (for RQ3): e.g., benchmarking, empirical study
    and SE tasks inspired by the IEEE's SWEBOK.
\end{itemize}

To answer RQ4, we have also recorded the following additional information:
\begin{itemize}
    \item Hardware: Which and how many devices were used to run the experiments.
   \item Measurement: How energy was measured (e.g., tool or estimation).
   \item Replication: How many repetitions were carried out and how were averages computed.
   \item Stability: How noise and overheads has been handled (e.g., warmup, cooldown).
   \item Dataset: What datasets are used in experiments.
\end{itemize}

To support reproducibility, we make the results from each stage of our search and classification process publicly available.\footnote{\url{https://figshare.com/s/3b85eb2f2a626ed81b1d?file=67470420}}

\subsection{Manual Labelling}
\label{sec-analysis}

We manually classified the collected studies
according to their application domain (RQ2) and research type (RQ3). To achieve this, two authors independently examined the titles of all included publications to assign them to relevant application domains (e.g., mobile, cloud, AI).
In cases of disagreement, the authors discussed their assessments and, when necessary, consulted the abstracts or full texts until consensus was achieved.
For RQ2, we observed that the two authors agreed with 81\% of labels.
An analogous procedure was applied to derive the subcategories related to the methods or approaches reported in the publications, with the two authors agreeing in 75\% of labels.\footnote{We rely on agreement scores given that we did not pre-define the set of categories beforehand.}

\subsection{Threats to Validity}
\label{section:threats}

{\bf Internal validity.}
Internal validity refers to
concerns with our methods that
could threaten the validity
of the results and claims made in our study.

First, a potential threat to internal validity
is the completeness of the reviewed literature, since
the risk of missing a publication cannot be eliminated.
To mitigate this issue,
we examine the publications of all top SE venues as mentioned in Section~\ref{section:search_procedure}
and define a set of inclusion and exclusion criteria presented in Section~\ref{section:selection_criteria}.

The reader should consider that, although our search (Table~\ref{table:venues}) included all top-rated software engineering (SE) conferences and journals, it did not cover every SE venue, nor did it involve a keyword-based search. Consequently, some relevant work may have been missed, particularly from venues such as the International Conference on Performance Engineering, the ACM Journal on Computing and Sustainable Societies, the International Conference on Requirements Engineering, and ICT4S (International Conference on Information and Communications Technology for Sustainability).
While publications from ICT4S were not included in our corpus (since it does not meet our current criteria, i.e., it is not listed as an A/A* SE venue), a future direction of this work would be to provide an overview of the applications studied in that venue and to examine the extent to which its findings align with or diverge from ours.
Despite these scope limitations of our survey,
we collected \papersRelevant publications,
which we believe are highly representative of Green SE
to provide a comprehensive overview of the field.

Second, another threat to the validity of survey refers to the analysis of the collected publications,
in particular their manual categorisation (RQ2, RQ3). 
The labelling process of publications with regards to their application field and type of study was conducted by two authors independently.
However, there could exist manual errors.
As such, we were able to discuss and resolve disagreements. 
We also provide an online repository, including our labelling decisions, to make this step reproducible.

Third, a possible threat to validity
refers to the filtering for RQ4
discussed in Section~\ref{filter-rq4}.
There could be more papers in the \papersRelevant
if we are more lenient and include more categories, fewer pages and so on. 
However, we argue that the filtering of the publications
was deterministic to get the subset of \papersExtract out of \papersRelevant
appropriate to answer RQ4.

{\bf External validity.}
External validity refers to the generalisability and reproducibility of the produced results and observations.
We provide a spreadsheet containing all the metadata for all the
publications selected in each of the phases
of our study.\footnote{\url{https://figshare.com/s/3b85eb2f2a626ed81b1d?file=67470420}\label{fn:shared}}

\section{Results}
\label{section:results}

\subsection{RQ1:Which top SE venues have served as publication targets for research addressing energy-related concerns?}

\begin{figure}[t]
\centering

\begin{subfigure}{0.57\columnwidth}
    \centering
    \includegraphics[width=\linewidth]{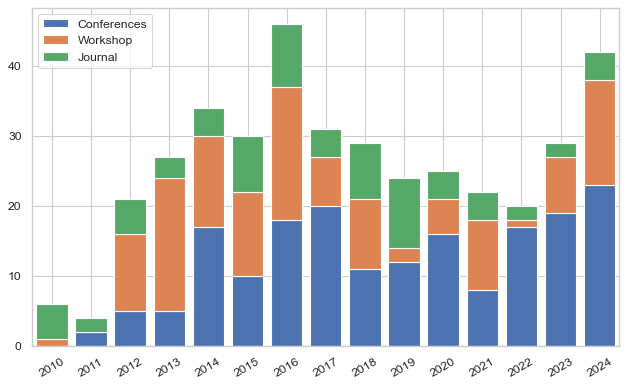}
    \caption{Number of publications per year. The darker the color the higher the number of publications.}
    \label{fig:years}
\end{subfigure}
\hfill
\begin{subfigure}{0.35\columnwidth}
    \centering
    \includegraphics[width=\linewidth]{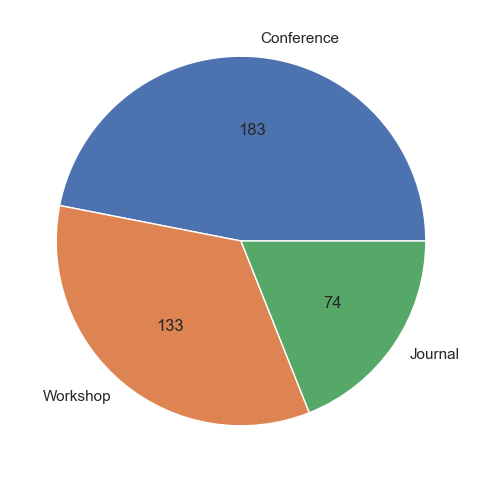}
    \caption{Number of publications per venue type}
    \label{fig:venuesPie}
\end{subfigure}

\caption{RQ1: Overview of Green SE publications over years (a) and per venue type (b) and per venue over years (c).}
\label{fig:combined}
\end{figure}

\begin{figure}[h]
\centering

\begin{subfigure}[T]{0.48\columnwidth}
    \centering
    \includegraphics[width=\linewidth]{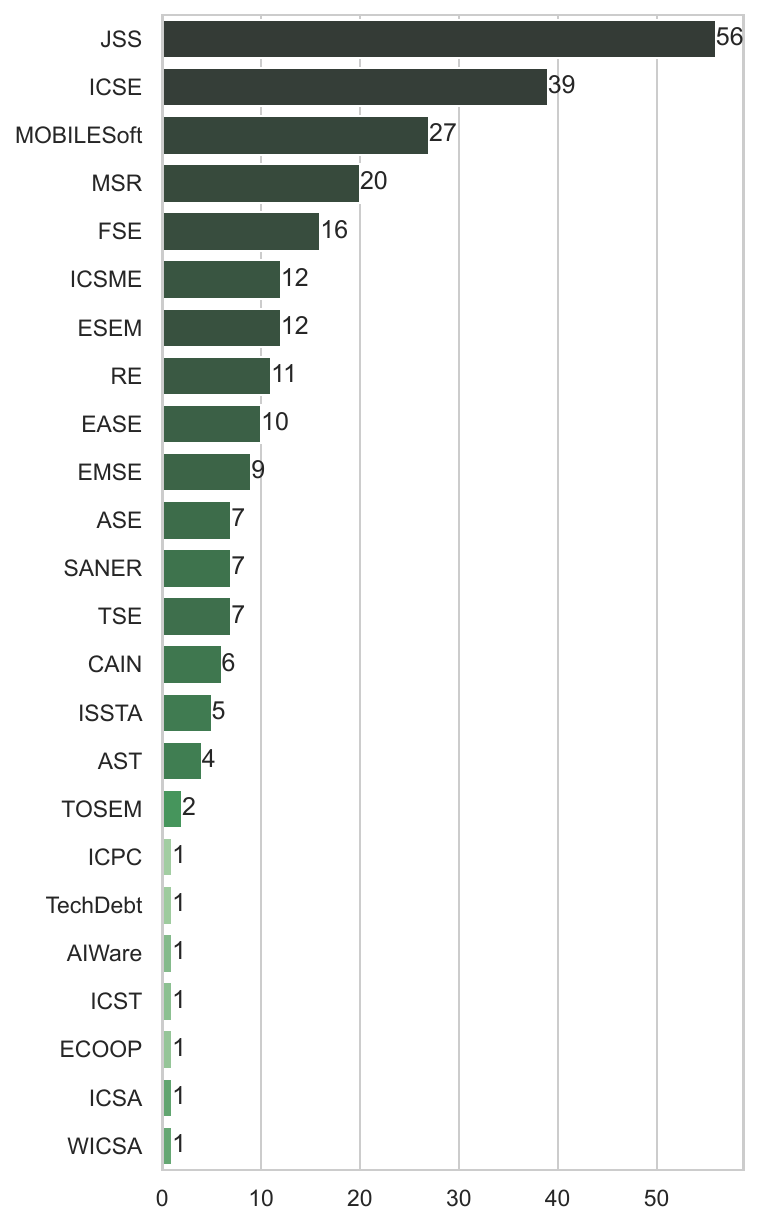}
    \caption{Conferences and Journals.}
    \label{fig:venuesConf}
\end{subfigure}
\hfill
\begin{subfigure}[T]{0.48\columnwidth}
    \centering
    \includegraphics[width=\linewidth]{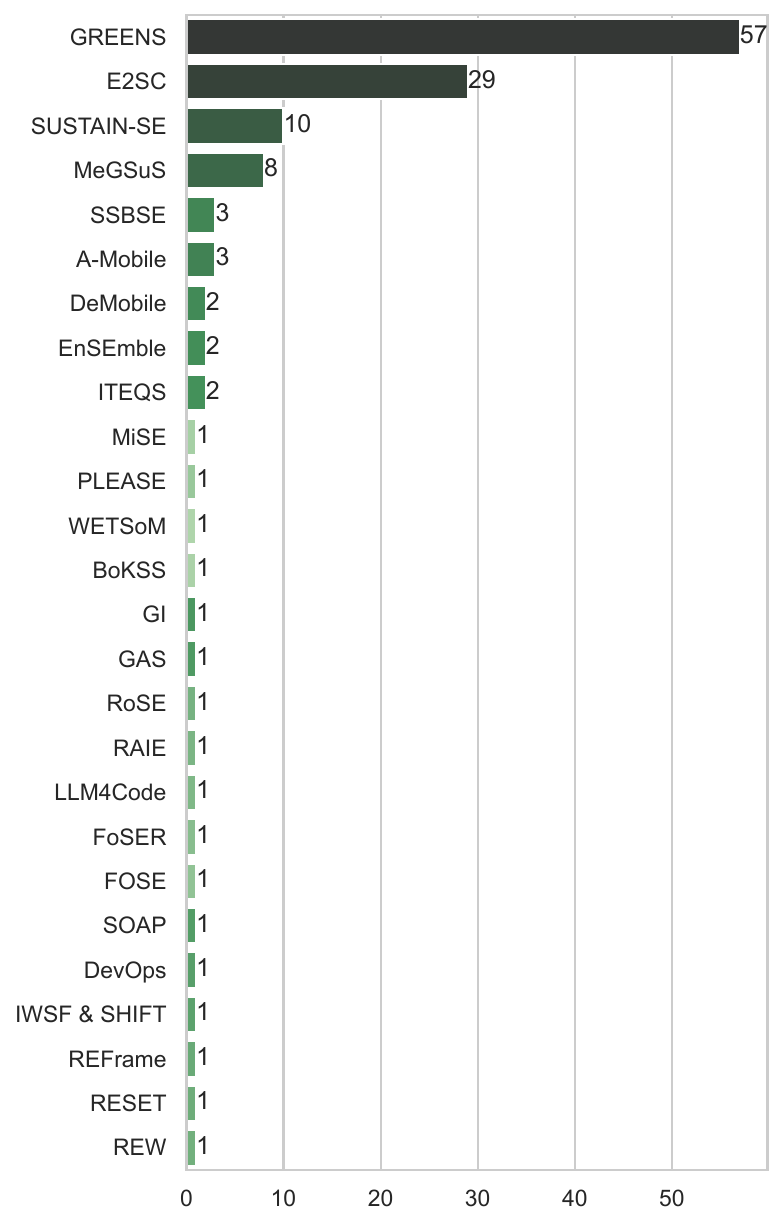}
    \caption{Workshops.}
    \label{fig:venuesWorkshop}
\end{subfigure}

\caption{RQ1: Total number of publications per venue.}
\label{fig:venues}
\end{figure}

RQ1 investigates the top SE venues (see Section~\ref{section:search_procedure})
in which energy-related SE research
has appeared over the past 15 years (2010-2024).

Figure~\ref{fig:years} summarises the trend of publications over time. We can observe that the number of publications peaks in 2016. Following this peak, the annual publication count declines until 2022,
after which it begins to rise again in 2023 and 2024, which --we conjecture-- is likely due to the booming of AI and the increasing interested in making it greener.

Figure~\ref{fig:venuesPie} shows the distribution of publications across different venue types, i.e., conferences, journals, and workshops. Conferences account for the largest share of publications (183 out of \papersRelevant), followed by workshops (133 out of \papersRelevant).
By contrast, journals contribute a relatively smaller number of publications, totalling 74 out of the \papersRelevant papers.

Figure~\ref{fig:venues} presents
the number of  publications per each of the venues considered. We can observe that GREENS and JSS account for the highest number of Green SE publications among the \numVenues distinct venues identified.
Notably, JSS is the only journal with more than ten relevant publications.
GREENS, which is dedicated to Green SE,
and, in 2026, it is going to hold its tenth edition, reflects the importance of the Green SE field.
The third highest-ranked venue is ICSE,
which has also hosted the GREENS workshop for several years.
Other confereneces with a substantial number of Green SE publications include MOBILESoft, and MSR, both co-located with the leading SE conference, ICSE.
Additional contributions can be found in other well-established top SE conferences including FSE (16 papers), ICSME (12 papers), and ESEM (12 papers).

Beyond GREENS, our analysis identifies three additional workshops that are explicitly dedicated to energy-related concerns and sustainable SE, which actively contribute to the dissemination of the Green SE research.
These workshops are: E2SC, SUSTAIN-SE, and MegSuS.
Another group of workshops promoting the Green SE research focuses on mobile applications, including A-Mobile and DeMobile. 

Although several venues host a considerable number of publications (e.g., nine venues with more than ten papers), this pattern is not consistent across all venues.
In particular, 24 out of the \numVenues venues include only a single relevant publication.
On the one hand, this distribution suggests that energy-related research is disseminated across a broad range of venues and reaches diverse audiences. On the other hand, such isolated publications may indicate that Green SE is not yet a primary thematic focus of top SE venues.

\subsection{RQ2: Which application domains are investigated the most with respect to energy-related concerns?}

\begin{table}[t]
\caption{RQ2: Application categories and their keywords/subtypes.}
\label{table:applications}
\centering
\rowcolors{1}{}{Gray2}
\scalebox{1.0}{
\begin{tabular}{ll}
\toprule
Category & Topics \\ \midrule
Software Engineering (SE) & Programming language, digital twin, software \\
Hardware & Hardware, HPC, Cluster, Containers, Grids, Data Center, Infrastructure \\
Mobile  & Wireless systems, Mobile, Embedded systems, Wireless networks \\
Artificial Intelligence (AI) & Artificial Intelligence, Machine Learning, Deep Learning, Large Language Models \\
Others & Quantum, Robotics, Blockchain, Virtual Reality \\
\bottomrule
\end{tabular}
}
\end{table}

\begin{figure}[t]
\includegraphics[width=.8\columnwidth]{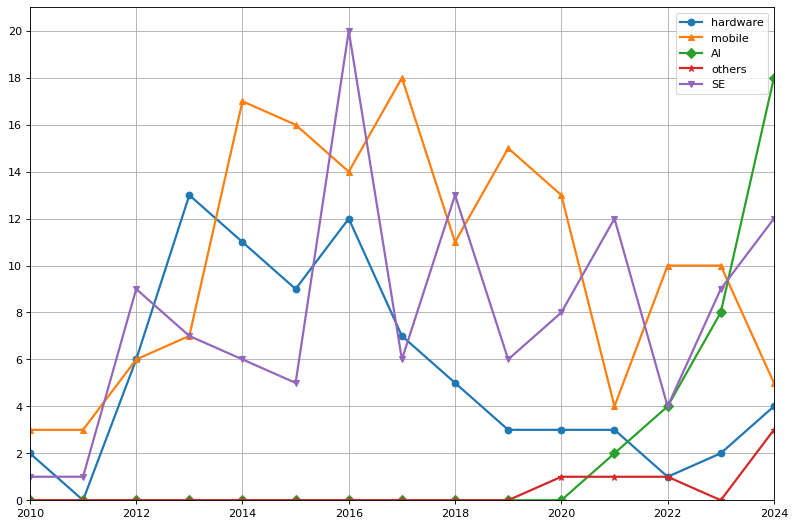}
\caption[]{RQ2: Number of papers per year for each of the application domains.}
\label{fig:rq2-years}
\end{figure}

Inspired by previous survey studies~\cite{verdecchia2023systematic,penzenstadler2014systematic} and based on the ACM’s Classification System,\footnote{https://dl.acm.org/ccs}, we defined five high-level categories to classify the application domains of the collected publications, namely AI, Hardware, Mobile, SE, Others. Table~\ref{table:applications} shows these category along with the topics associated with each.
Among them, the Mobile domain accounts for the largest number of publications (152),
followed by SE with 119 publications. Hardware- and AI-related studies follow, with 81 and 32 publications, respectively.
The category labelled ``Others'' comprises publications that could not be assigned to any of the four primary categories and includes only six papers. Figure~\ref{fig:rq2-years} shows the number of papers per year (2010–2024) across five application domains: Hardware, Mobile, AI, Others, and SE. A striking pattern emerges:
the publication activity is domain-shifted over time rather than uniformly growing.
Early years (2010–2013) are dominated by hardware and SE. Mid period (2014–2019) is driven primarily by mobile and SE.
Recent years (2021–2024) show a strong surge in AI, which becomes the leading domain by the end of the timeline. The ``others'' category remains consistently minimal throughout. 

The Hardware domain shows early growth, rising from low counts in 2010–2012 to a peak in 2013 ($\approx$13 papers). After 2016, there is a steady decline, reaching low single digits from 2019 onwards. Only a slight recovery appears after 2022. These observations suggest that hardware research concerning energy-related aspects  was initially prominent, but gradually lost relative research attention. We can speculate that this is due to a very fast advance in hardware capabilities in early years, which is less significant nowadays.

The Mobile domain has assisted to a rapid increase in publications number from 2012, becoming the dominant domain between 2014–2018, with a peak in 2017 (18 papers). Afterward, we observe various fluctuations with a noticeable dip in 2021 (from $\approx$14 to 4 articles), a partial increase (10 articles) in 2022–2023, and another small drop in 2024.
These observations suggest that mobile research concerning energy-related aspects experienced a strong mid-period wave, then stabilised around 4-10 papers per year.

As for the AI domain, there is near-zero activity until 2020, indicating minimal early focus on energy-related concerns in this domain. A sharp growth begins around 2021, accelerating through 2023–2024, becoming the top domain by 2024 (18 papers).
AI energy-related concerns is a late but rapidly expanding research focus, showing the strongest recent momentum.

While the domains discussed so far show variable trends over time, the SE domain, is present throughout the entire period with consistent output. It shows a major spike in 2016 (20 papers) — the only highest yearly value across all domains (which also corresponds to the year with the highest overall number of publications, see Figure~\ref{fig:years}).
Afterward, SE maintains moderate but steady counts (roughly 4–12 papers/year).
These results suggest that investigating SE energy-related concerns is a stable, foundational domain with one notable surge in mid-period.

The Others domain are scarcely represented across all years, with small up-ticks appearing only after 2020, but never departing from very low number of publications, thus suggesting that these topics remain peripheral compared to the main domains.

We conclude that there has been a clear domain shift over time: from Hardware and SE to Mobile to AI. AI energy-related research shows the strongest and most recent growth trajectory, suggesting it is currently the primary driver of new publications in this research area, while Hardware energy-related research has become comparatively less emphasised and SE energy-related research remains consistently relevant.

\subsection{RQ3:What types of research approaches are exploited to tackle energy-related concerns in different SE tasks?}

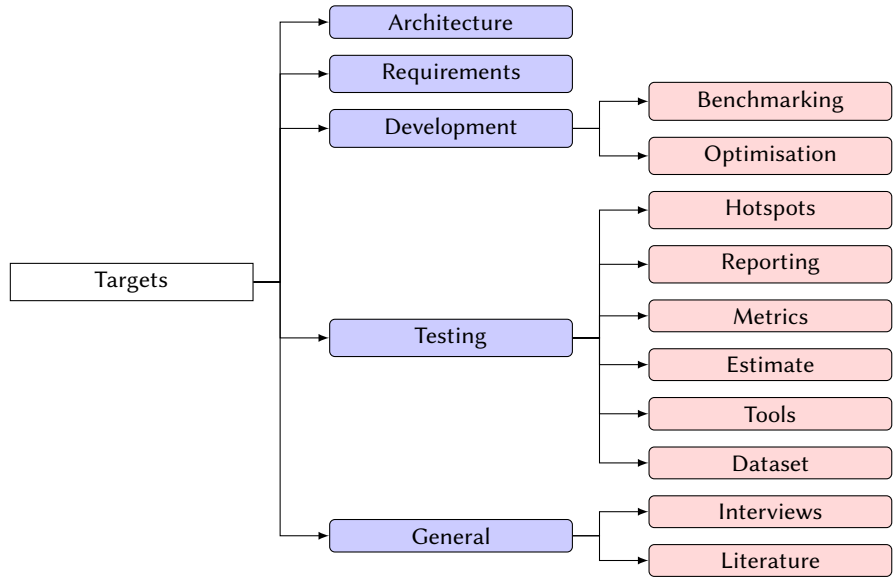
\begin{figure}
\tikzset{
    basic/.style  = {draw, text width=3cm, align=center, font=\sffamily, rectangle},
    root/.style   = {basic, rounded corners=2pt, thin, align=center, fill=green!30},
    onode/.style = {basic, thin, rounded corners=2pt, align=center, fill=green!60,text width=3cm,},
    tnode/.style = {basic, thin, align=left, fill=pink!60, text width=10em, align=center},
    xnode/.style = {basic, thin, rounded corners=2pt, align=center, fill=blue!20,text width=3cm,},
    xnode2/.style = {basic, thin, rounded corners=2pt, align=center, fill=pink!60,text width=3cm,},
    wnode/.style = {basic, thin, align=left, fill=pink!10!blue!80!red!10, text width=6.5em},
    edge from parent/.style={draw=black, edge from parent fork right}

}

\begin{forest} for tree={
    grow=east,
    growth parent anchor=west,
    parent anchor=east,
    child anchor=west,
    edge path={\noexpand\path[\forestoption{edge},->, >={latex}] 
         (!u.parent anchor) -- +(10pt,0pt) |-  (.child anchor) 
         \forestoption{edge label};}
}
[Targets, basic,  l sep=10mm,
    [General, xnode,  l sep=10mm,
        [Literature, xnode2,  l sep=10mm]
        [Interviews, xnode2,  l sep=10mm]]
    [Testing, xnode,  l sep=10mm,
        [Dataset, xnode2,  l sep=10mm]
        [Tools, xnode2,  l sep=10mm]
        [Estimate, xnode2,  l sep=10mm]
        [Metrics, xnode2,  l sep=10mm]
        [Reporting, xnode2,  l sep=10mm]
        [Hotspots, xnode2,  l sep=10mm]]
    [Development, xnode,  l sep=10mm,
        [Optimisation, xnode2,  l sep=10mm]
        [Benchmarking, xnode2,  l sep=10mm]]
    [Requirements , xnode,  l sep=10mm]
    [Architecture , xnode,  l sep=10mm]
         ] 
\end{forest}

    \caption{RQ3: Overview of study types and the respective publications.}
    \label{fig:tree}
\end{figure}

\begin{figure}[t]
\centering
\includegraphics[width=1\columnwidth]{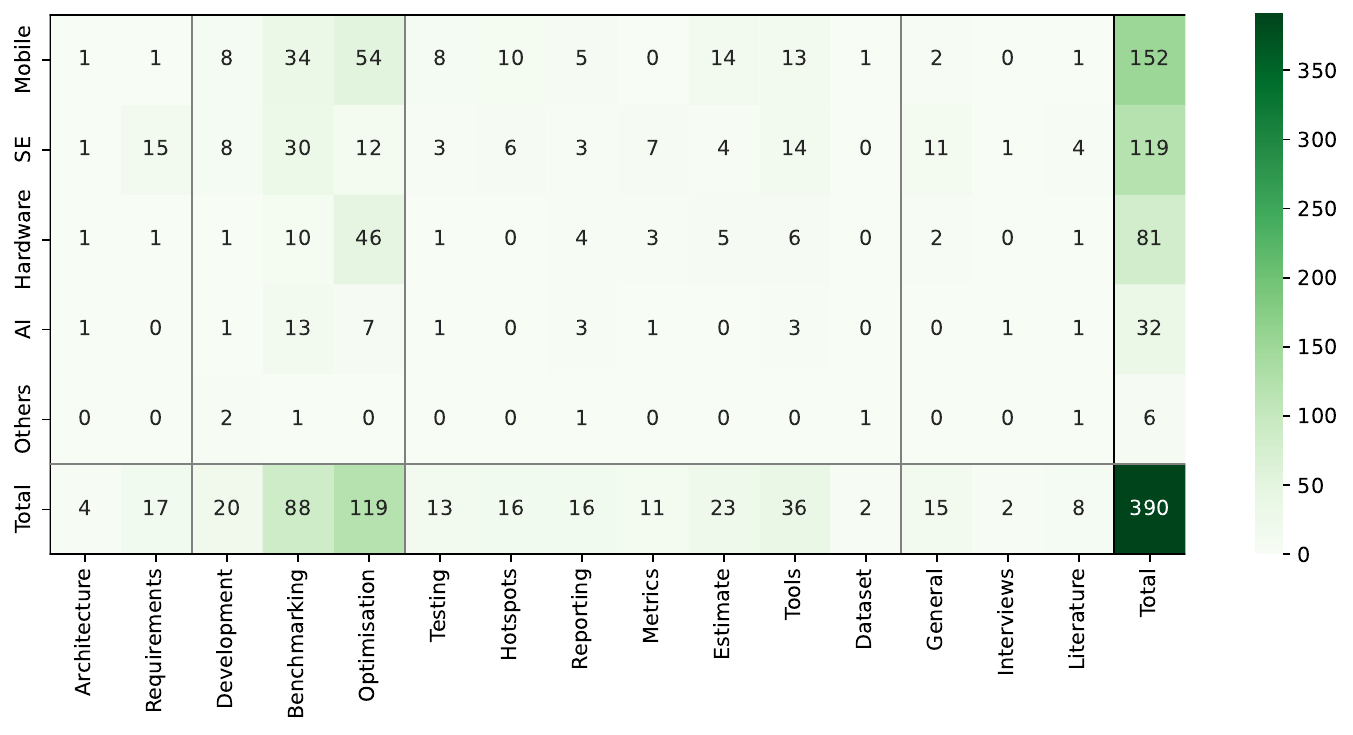}
\caption[]{RQ3: Heatmap for both categories combined.}
\label{fig:heatmapCategories2}
\end{figure}

Following the classification of the \papersRelevant publications by application domain and considering the IEEE's SWEBOK,\footnote{https://www.computer.org/education/bodies-of-knowledge/software-engineering} we conduct a second categorisation to identify the research focus within the SE domain as well as the research methodologies employed.

To this end, we define five categories, as illustrated in Figure~\ref{fig:tree}. Four categories correspond to stages of the software development lifecycle: Architecture, Requirements, Development, and Testing.
``Architecture'' refers to software design and ``Requirements''
to the functional and non-functional requirements
of the software under development.
``Development'' primarily comprises two types of studies, i.e., \textit{benchmarking} and \textit{optimisation}, which support software engineers during implementation, for instance by recommending energy-efficient data structures or algorithmic choices.
``Testing'' encompasses five types of studies, including the identification of energy \textit{hotspots}, the \textit{reporting} of energy consumption, and the definition and application of energy-related \textit{metrics}. In addition, this category includes research on energy measurement \textit{tools} and \textit{estimation} techniques.
The ``General'' category includes publications with a broader perspective on Green SE that cannot be clearly assigned to any specific lifecycle stage (e.g., Architecture, Requirements, Development, or Testing).
As such, literature reviews and interview studies are included in the ``General'' category.

Figure~\ref{fig:heatmapCategories2} summarises the distribution of publications across the categories defined in Figure~\ref{fig:tree}, together with the application domains introduced in RQ2. The Development stage emerges as the most extensively studied research area, with a substantial number of publications focusing on benchmarking (88 publications) and optimisation (119 publications). Optimisation studies are particularly prevalent in the Mobile and Hardware application domains. All five subcategories within the Testing stage are represented by at least ten publications. By contrast, Architecture and Requirements are comparatively unexploded in the analysed corpus. 
Particularly, only four publications addressed Architecture while 17 considered Requirements.
We also identify eight literature reviews and two interview studies.

\subsection{RQ4: What are the key elements of the experimental design protocols considered in SE studies that address energy-related concerns?}

To address RQ4,
we study the experimental design protocols
reported in the \papersExtract studies identified from the selection process
mentioned in Section~\ref{selection_process}.
The analysis highlights a set of recurring design elements
that characterise how empirical evaluations have been performed
in the studies examined.
These elements can be organised into four main dimensions:
(i) hardware selection and configuration;
(ii) execution environment stability;
(iii) repetition strategies;
and (iv) energy measurement tools and techniques.
The following paragraphs describe these dimensions in detail.

{\bf Hardware.}
Hardware selection and configuration refer to a critical component of the experimental design in the examined studies.
Many works employ multiple and heterogeneous devices to enable comparative evaluations (e.g., \cite{couto2020energy,chowdhury2018exploratory,10.1109/MOBILESoft.2017.7,DBLP:journals/jss/SinghTS24,rua2020greenspecting,sahin2016benchmarks}). Specifically, studies involving AI techniques frequently rely on specialised hardware, such as GPUs, reflecting the computational demands of these workloads.
Several experimental protocols further account for device diversity not only in terms of model or architecture (e.g., \cite{georgiou2018your, DBLP:conf/msr/OliveiraOC17, DBLP:conf/mobilesoft/JacquesAC24, DBLP:conf/msr/OliveiraOCF019, oliveira2021improving}), but also device age. Consider that battery ageing has been shown to affect energy consumption (e.g., \cite{bangash2021energy}).
In addition to execution platforms, some studies explicitly specify the use of dedicated measurement hardware,
with higher-precision devices typically associated with higher acquisition costs~\cite{bruce2018approximate}.

\begin{table}[t]
\caption{RQ4: Measurement tools.}
\label{table:tools}
\centering
\rowcolors{1}{}{Gray2}
\begin{tabular}{ll}
\toprule
Measurement tool & References \\ \midrule
RAPL & \cite{DBLP:journals/jss/LimaSLCMF19,DBLP:conf/icse/0001ZK0KB24,10.1145/3475716.3475774,lima2016haskell,melfe2018helping,de2021runtime,georgiou2022green,DBLP:conf/icsm/OurnaniRRP21} \\
Monsoon power monitor &\cite{DBLP:conf/icse/NikzadCG14,10.1145/3387905.3392095,10.1145/2884781.2884867,10.1145/2593743.2593750,10.1145/3241742,DBLP:conf/icse/LiTH14,DBLP:conf/icse/BanerjeeR16,10.1145/2897073.2897085,hampau2022empirical} \\
jRapl& \cite{DBLP:conf/kbse/MacedoAGPS20,DBLP:conf/esem/RochaC019,DBLP:conf/sigsoft/BabakolCMSL20,DBLP:conf/icsm/PintoLCL16,DBLP:conf/msr/OliveiraOCF019,DBLP:conf/greens/PereiraCSCF16,oliveira2021improving} \\
Watts up? PRO& \cite{DBLP:journals/ese/BreeC23,georgiou2018your,DBLP:conf/esem/ProcacciantiLD16,bree2022removing,procaccianti2016empirical} \\
BatteryManager & \cite{DBLP:conf/mobilesoft/ChenLCZTKM24,DBLP:conf/sigsoft/CaninoLM18,DBLP:conf/greens/NevesLNPEIM23,bogdan2024empirical} \\
GreenMiner& \cite{DBLP:journals/ese/McIntoshHH19,chowdhury2016client,DBLP:conf/icse/HasanKHSAH16,chowdhury2018exploratory} \\
pyRAPL & \cite{DBLP:conf/rose-ws/AlbonicoVRW24,DBLP:conf/icsa/TedlaKV24,brownlee2021exploring} \\
trepn & \cite{couto2020energy,10.1109/MOBILESoft.2017.7,rua2020greenspecting,janssen2022impact,van2022comparing} \\
INA219 energy measurement chip & \cite{DBLP:conf/icse/HasanKHSAH16,10.1145/2593743.2593749,rasmussen2014green} \\
Android Power Profiler & \cite{DBLP:conf/msr/OliveiraOCF019,oliveira2021improving} \\
National Instruments USB-6215 & \cite{10.1145/2652524.2652538,manotas2014seeds} \\
PowerJoular & \cite{wagner2023energy,nahrstedt2024empirical} \\ 
Android SDK & \cite{DBLP:journals/jss/SinghTS24,DBLP:conf/msr/OliveiraOC17} \\
Batterystats & \cite{wattenbach2022you} \\ 
Ebserver & \cite{oliveira2023analyzing} \\
Pyjoules & \cite{DBLP:conf/msr/ShanbhagC23} \\
MSPSim &  \cite{DBLP:conf/ecoop/LiW23} \\
EET &  \cite{DBLP:journals/jss/FernandezCMSG24} \\
Fluke©i30 AC/DC Current Clamp &  \cite{DBLP:conf/icse/BartensteinL13} \\
iGreenMiner &  \cite{bangash2021energy} \\
Power sensors &  \cite{DBLP:conf/icse/CruzA17} \\
dumpsys10 &  \cite{malavolta2020evaluating} \\
NAGA VIPER ammeter &  \cite{carette2017investigating} \\
NI LabVIEW SignalExpress &  \cite{DBLP:conf/icsm/PintoLCL16} \\
USB Tester OLED Backpack 2.0 1 &  \cite{DBLP:conf/greens/RashidAT15} \\
NVIDIA System Management Interface &  \cite{DBLP:conf/cain/Yarally0FSD23} \\
MAGEEC Energy Measurement &  \cite{bruce2018approximate} \\
Yokogawa WT210 &  \cite{7889026} \\
Spartan-6’s power monitors &  \cite{sahin2012initial} \\
emonTx V3 energy monitoring node &  \cite{8118155} \\
GW Instek GPM-8213 digital power meter &  \cite{DBLP:conf/kbse/TundoMIBBM23} \\
TiePie Handyscope HS5 &  \cite{8052533} \\ \bottomrule
\end{tabular}
\end{table}

{\bf Measurement.}
The choice of energy measurement approach constitutes a key element of the experimental design of Green SE studies, as it determines measurement granularity and accuracy (e.g., sampling rate \cite{georgiou2018your, DBLP:conf/msr/OliveiraOCF019, 8118155}). The surveyed studies exhibit substantial diversity in tool selection, with no dominant standard emerging. Within the benchmarking and optimisation studies (RQ3), we identify 31 distinct tools used to measure or estimate energy consumption (Table~\ref{table:tools}).

While many tools appear only sporadically, two approaches are comparatively common: RAPL-based solutions and the Monsoon power monitor. RAPL is used either directly or via wrappers, such as jRAPL and pyRAPL, and is often accessed through higher-level tools (e.g., perf, PyJoules). The Monsoon power monitor, an external watt meter, is predominantly used in mobile application studies, whereas other external meters (e.g., Watts Up? PRO) are mainly adopted in general-purpose software evaluations. Hardware-based measurement approaches are frequently preferred due to their non-intrusive nature and lack of execution overhead (e.g., \cite{sahin2012initial, DBLP:conf/icsm/SahinTMPC14, sahin2016benchmarks}).

Several studies also incorporate estimation-based techniques, particularly for mobile and machine learning workloads. For mobile applications, energy consumption may be inferred from parameters such as screen brightness and battery reporting granularity (e.g., \cite{DBLP:conf/sigsoft/VasquezBBOPP15, 10.1145/3241742}). For machine learning workloads, dedicated calculators, such as the Machine Learning Emissions Calculator (e.g., \cite{shi2024greening}), are often used to estimate energy usage.

{\bf Stability.}
A common design practice is the explicit control of the execution environment to reduce noise and confounding effects.
Most studies report disabling background processes prior to experimentation.
For mobile platforms, additional environmental parameters are frequently controlled, including screen brightness (e.g., \cite{DBLP:conf/greens/NevesLNPEIM23, DBLP:conf/mobilesoft/ChenLCZTKM24, DBLP:conf/sigsoft/CaninoLM18, 10.1145/3387905.3392095, carette2017investigating, couto2020energy, DBLP:conf/icse/HasanKHSAH16, bangash2021energy, bogdan2024empirical, 10.1145/3241742}) and battery charge level (e.g., \cite{DBLP:conf/sigsoft/CaninoLM18, carette2017investigating, DBLP:conf/msr/OliveiraOC17, DBLP:conf/mobilesoft/JacquesAC24, DBLP:journals/jss/SinghTS24}).
Warm-up executions are also commonly incorporated into experimental protocols (e.g., \cite{couto2020energy, DBLP:conf/msr/OliveiraOCF019, DBLP:conf/greens/PereiraCSCF16, oliveira2021improving,oliveira2023analyzing,hampau2022empirical}) to address dependency loading and thermal stabilisation. 
Furthermore, cool-down intervals between consecutive runs are often specified, with durations ranging from a few seconds (e.g., \cite{DBLP:conf/mobilesoft/ChenLCZTKM24, DBLP:conf/icse/BartensteinL13, couto2020energy, de2021runtime}) to several minutes (e.g., \cite{DBLP:conf/greens/NevesLNPEIM23, georgiou2018your, carette2017investigating}).

\begin{figure}[t]
\includegraphics[width=.6\columnwidth]{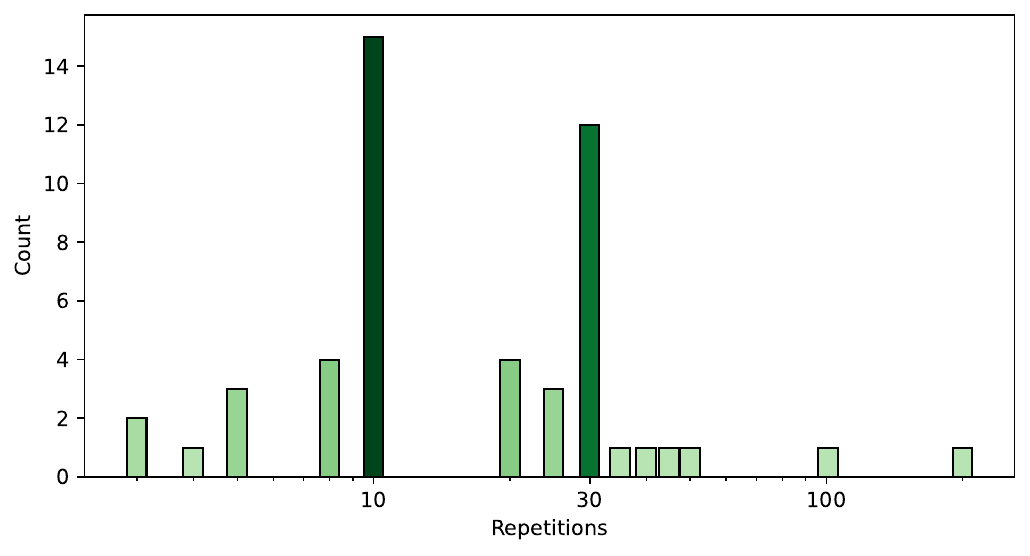}
\caption[]{RQ4: Number of repetitions for experiments per paper.}
\label{fig:rq4-reps}
\end{figure}

{\bf Repetition.}
Repetition is widely used as a design mechanism to improve generalisability and mitigate measurement variability.
However, the number of repetitions is often constrained by the computational cost of the experimental workload.
This limitation is particularly evident in studies involving deep learning and large language models, where training and evaluation can span over weeks or months, or are simply too costly~\cite{williams2026reflection}.
Consequently, experimental protocols frequently reflect a trade-off between statistical reliability/accuracy and practical feasibility, influencing the extent to which repetitions can be performed.
From Figure~\ref{fig:rq4-reps}, we observe that the most frequently choice for number of repetitions are 10 and 30, with the maximum being at 200.

\textbf{Datasets.}
Our analysis reveals that
the datasets used in the empirical studies
under examination differ according to
the application domain of each study.
Namely, there are no global datasets used in all the different studies.
For each application domain, however,
there are some common benchmarks and practices considered.

Specifically, we found out that for measuring
the energy consumption of programs written in different programming languages,
the empirical studies mostly consider the following benchmarks:
the Computer Language Benchmarks Game (6 studies),\footnote{https://benchmarksgame-team.pages.debian.net/benchmarksgame/index.html}
the Rosetta Code (5 studies),\footnote{https://rosettacode.org/wiki/Rosetta\_Code} and
the DaCapo (5 studies)~\cite{blackburn2006dacapo}.
When the empirical studies measure the energy consumption of
machine learning,
the use of the UCI archive\footnote{https://archive.ics.uci.edu/} is common,
while for deep learning algorithms the use of the LEARNINGEXAMPLES\footnote{https://github.com/nvidia/deeplearningexamples} from NVIDIA is also preferable.
As expected, for studies that measure energy consumption
of mobile apps,
Android apps, mostly from F-Droid\footnote{https://f-droid.org/en/}
are used.
Furthermore, several studies use specific benchmarks
that could be also more largely considered
such as the GreenHub~\cite{MCC19}, CUAD~\cite{DCA21}, PARSEC~\cite{Bie2011}, Devign, BigCloneBench, and so on.
Finally, studies that measure the energy consumption of web apps choose their own a few apps to test, such as in~\cite{DBLP:conf/greens/NevesLNPEIM23}.

In general, for all types of studies,
the number of subjects is limited.
This may happens become of the time required for the experiments
to measure energy and for hardware requirements, as well as costs.

\section{Implications}
\label{section:implications}
A great deal of work has mentioned research and practical implications
regarding green software~\cite{LM26,CFK25,MJ24}.
Here, based on our findings from the research questions presented in Section~\ref{section:results},
we derive the following research and practical implications for Green SE.

\subsection{Green SE research as a first-class citizen}

Our findings align with the recent study by Williams et al.~\cite{williams2026reflection}, arguing that having energy-related results in research papers
has been considered a ``nice-to-have''.
This mindset must flip to enable the so-called ``Green Shift Left''~\cite{MJ24}.
To do so, the SE community can take several actions as follows.

{\bf Venues.} SE peer-reviewed journals and conferences
could encourage reviewers and editors
to systematically inquire about any energy measurements when evaluating research papers,
just as they do with open science.
This would reveal the necessity of green software,
especially in an era where AI-powered systems consume significant amounts of energy.
Furthermore, having more dedicated conferences, workshops, such as GREENS,\footnote{Note that if there is an increasing interest of the community, then workshops such as GREENS may be upgraded to conferences.}
invited talks, and data challenge tracks, such as the MSR challenge track (i.e., a collaborative research competition, where participants analyze a designated software repository dataset to answer fresh research questions.),\footnote{https://2026.msrconf.org/track/msr-2026-mining-challenge?\#Call-for-Mining-Challenge-Papers}
would further elevate energy-aware SE research
within the main tracks of leading venues
and reinforce the importance of energy considerations in SE.
According to Figure~\ref{fig:venues}, nowadays, there are only a few small dedicated workshops on sustainability (GREENS, E2SC, SUSTAIN-SE, MeGSuS, BoKSS).
Another way to support dissemination is the inclusion of invited talks to increase the number of energy related presentations at general venues (i.e., venues where sustainability is not a focus).
For instance, this has been done at the following workshops: DeMobile'14, DeMobile'15, WODA'14 and WODA+PERTEA'14.
Finally, there should be venues (e.g., seminars, tutorials, summer schools) towards educating researchers about sustainability and energy concerns.
For instance, consider the IASESE Advanced School ``Empirical studies targeting green and sustainable software and energy measurement'' conducted at ESEM in 2024.
The one-day event included topics such as the different dimensions of software sustainability and the carrying out of empirical studies.

{\bf Synergy.} Furthermore, publishing results from field experiments
using real-world data will emphasise
the practical impact of Green SE and further highlight its importance.
In line with our findings for RQ4,
advancing the state-of-the-art requires
new large-scale real-world datasets and empirical studies.
To achieve this goal,
academics and practitioners
must establish a systematic collaboration.
Finally, initiatives, such as the
Green Software Foundation\footnote{https://greensoftware.foundation/}
may bring together practitioners and researchers
seeking to effectively optimize energy consumption.

{\bf Reproducibility.}
There should be a systematic way
to conduct experiments on Green SE.
To do so, clear guidelines, standardised benchmarks, and open science
could help~\cite{LM26}.
Our RQ4 lists the key criteria for
such experiments.
Furthermore, the use of containers such as Docker could make easier the execution of the experiments.
Finally, since such experiments may require expensive hardware and real-world data,
research funding and collaboration with the industry
could also assist.

\subsection{Real-world deployment}
According to Heldal et al.~\cite{HNM24},
while there is a growing number of academic efforts towards sustainability,
there is not yet comprehensive research on the competencies and skills required
by IT professionals to develop such sustainable systems.
Therefore, such systems are not broadly available and used in industrial settings.
The industry needs to be more energy aware.
In the following paragraphs,
we discuss our findings and relevant suggestions
relevant to the concerns of Heldal et al.~\cite{HNM24}.

{\bf Environmental consciousness across SE application domains.}
According to our findings in RQ2,
mostly the mobile domain refers to energy-related concerns.
In AI for SE, researchers have merely scratched the surface on how to measure and report energy \cite{williams2026reflection}.
This should be considered as a red flag for our community
since energy demands on AI experiments are dramatically increasing~\cite{MJ24}.
 More systematic studies are needed in different
SE fields.
To do so,
more datasets should be open-source
and clear guidelines should be given
for experiments to be conducted on different fields.
Furthermore, dedicated tooling and hardware should
be used.
The combination of different fields
can be helpful here
since approaches used in one field can be used in another and vice versa.

{\bf Tooling for green-aware software development.}
Researchers require easy-to-use tools
for measuring energy at a minimum cost.
There is a need for the development of such tools e.g., {\tt FECoM}~\cite{RWS24} within industrial settings.
To do so, there should a synergy between academics and practitioners.
The tools should be usable, reliable and scalable.
As suggested by the related work~\cite{MJ24,LM26},
there is a need for developing tools that measure energy consumption
and report that back to the users to make them more energy aware.
Specifically, significant industrial effort has been focused on minimizing the costs of IT infrastructure, such as datacenters that require significant power and cool environments to operate~\cite{MJ24}.

\section{Related Work}
\label{section:rw}

In this section, we present surveys that are most closely related to our work.
We consider both surveys on Green SE (including those related to mobile and cloud computing) and
surveys on Green AI.
We include Green AI surveys because, as discussed in RQ2, Green AI has
emerged as a significant application domain within Green SE.
Given the rapid growth of AI, Green AI constitutes a relevant
subfield for understanding recent developments in Green SE.
Finally, we compare the scope and findings of our survey with
those of existing related surveys.

\subsection{Surveys on Green SE}
\label{greense}
Several surveys have been conducted addressing software sustainability and energy-related concerns focusing either on software development processes or on a specific area (e.g., datacenters~\cite{kong2014survey}, cloud computing~\cite{mastelic2014cloud}, mobile devices~\cite{hoque2015modeling}, processors~\cite{mittal2014survey}, database systems~\cite{guo2022energy}, metrics for assessing green software~\cite{bozzelli2013systematic}).
Here, we present the closest works to our study,
providing an overview of the literature on Green SE. 

Mour{\~a}o et al.~\cite{mourao2018green} performed a mapping study on green and sustainable software engineering, which considered Green SE and included \papersExtract studies up to 2017, two of which are prior to 2010 (2003 and 2009).
They investigated whether sustainability is considered in the software development lifecycle. They report on the details of each study's design: the research method (e.g., case study, survey), research type (e.g., solutions, evaluation study), contribution type (e.g., framework, tool, metrics), type of evidence (e.g., simulations). 
Other investigated aspects include the application domains, venue and amount of industry and academia contributions.

We have encountered several surveys and mapping studies conducting their search prior to 2018~\cite{berntsen2017sustainability,penzenstadler2014systematic,marimuthu2017software,wolfram2017sustainability}.
Among these, the most relevant to our study are the mapping studies by Penzenstadler et al.~\cite{penzenstadler2014systematic} and Marimuthu and Chandrasekaran~\cite{marimuthu2017software}.
Penzenstadler et al.~\cite{penzenstadler2014systematic} conducted a mapping study covering sustainable software from two perspectives: software being sustainable, and software supporting sustainability. 
Their study included 83 studies up to 2013, with the oldest being from 1989.
These works are analysed with regard to research topics, knowledge areas, research types (e.g., exploratory or opinion studies) and are divided in 10 application domains. 
Marimuthu and Chandrasekaran~\cite{marimuthu2017software} carried out a mapping study including 82 studies from 2010 to May 2016 on green and sustainable software, for which they considered a total of seven research questions. 
Among others, they considered the research goal (developing green software Vs. green software development), tool sharing and publication venues. 

A recent survey by Lee et al.~\cite{lee2024survey} collected 101 papers from 2014 to 2021, extracting information about Green SE from publication databases. 
Publications are categorised along the Software Development Life Cycle and divided in ten areas based on the Software Engineering Body of Knowledge (SWEBOK).
They also provided an overview of tools, which can support developers in creating energy-efficient software, and summarised several challenges of Green SE. 
Moreover, they suggest a brief list of solutions to support practitioners with green SE practices.

\subsection{Surveys on Green AI}
\label{greenai}

Beyond existing surveys on Green SE,
work from the Green AI literature should also be considered,
as this field is rapidly emerging.
Software engineering solutions may incorporate machine learning (ML) components,
which are known to be energy intensive.
This highlights the importance of studying ML models with respect to the
energy consumption and efficiency of software systems.

For instance, Verdecchia et al.~\cite{verdecchia2023systematic}
reviewed the field of Green AI,
which focuses on reducing the carbon emissions of AI models.
Their study provided an overview of 98 papers published between 2015 and 2022.
From these studies, they extracted several types of information,
including venue type (journal, conference, workshop),
application domain (e.g., edge, computer vision, cloud, mobile),
the phase of the AI model,
the studied artifact used to achieve energy reductions,
and data characteristics (e.g., data type and dataset size).
Xu et al.~\cite{xu2021survey}
presented a survey of approaches for improving the energy efficiency of
deep learning  (DL) models.
These approaches are organised according to architecture,
training, inference, and data usage.
Rolnick et al.~\cite{rolnick2022tackling}
offered a complementary perspective on Green AI.
Rather than focusing on how AI systems can be energy efficient,
they discussed how AI can be used to reduce greenhouse gas emissions.

\subsection{Comparison}
Unlike existing surveys,
we conduct a systematic mapping study of Green SE (finding references on Green AI as a subfield) publications from 2010 to 2024,
with a particular focus on publication venues, application domains,
software development life cycle (SDLC) phases, and best practices.
Although previous surveys have reported venue-related information
(e.g., the number of papers published in journals Vs. conferences),
our study places greater emphasis on the application domains addressed by the literature, designing our search methods accordingly (Section~\ref{section:method}).

Given the parallels with prior studies,
we compare our findings with theirs to show if and how trends changed with regards to SE studies from prior years (i.e., 2014 \cite{penzenstadler2014systematic}, 2017 \cite{marimuthu2017software}, 2018 \cite{mourao2018green}, 2024 (but papers up to 2021) \cite{lee2024survey}),
as well as with trends in Green AI~\cite{verdecchia2023systematic}.
Finally, we make all survey results publicly available
to support open science and facilitate replication where possible.

Table~\ref{tab:green_se_green_ai_surveys}
provides a comparative summary of surveys most closely related to our work
(the last row corresponds to this study).
Specifically, we compare findings across four dimensions:
evolution and publication trends (RQ1),
application domains (RQ2),
software development life-cycle phases, according to SWEBOK (RQ3),
and common current empirical practices (RQ4).
The following paragraphs discuss these aspects in detail.

\begin{table}[htbp]
\centering
\footnotesize
\begin{tabular}{l l l c l c c c c}
\toprule
\textbf{Survey} & \textbf{Scope} & \textbf{Years} & \textbf{\# Studies} &
\textbf{Method} & \textbf{Venues} &
\textbf{Domains} & \textbf{DLC phases} &
\textbf{Practices} \\
 & & & & & \textbf{(RQ1)} & \textbf{(RQ2)} &
\textbf{(RQ3)} & \textbf{(RQ4)} \\
\midrule
\cite{mourao2018green}              & Green SE            & before 2018 & 75  & SMS      & YES & YES & YES & NO \\
\cite{berntsen2017sustainability}   & Green SE            & 2010--2016  & 36  & SMS      & YES & NO  & NO  & NO \\
\cite{penzenstadler2014systematic}  & Green SE            & 2012--2014  & 83  & SMS      & NO  & YES & YES & NO \\
\cite{marimuthu2017software}        & Green SE            & 2010--2016  & 82  & SMS      & YES & NO  & YES & NO \\
\cite{wolfram2017sustainability}    & Green SE            & 1980--2013  & 168 & SMS      & NO  & NO  & YES & NO \\
\cite{lee2024survey}                & Green SE            & 2010--2021  & 101 & SLR      & NO  & NO  & YES & YES \\
\cite{verdecchia2023systematic}     & Green AI            & 2015--2022  & 98  & SLR      & YES & YES & NO  & NO \\
\cite{xu2021survey}                 & Green AI            & before 2021 & NA  & SLR      & NO  & NO  & NO  & YES \\
\cite{rolnick2022tackling}          & Green AI            & before 2022 & NA  & Overview & NO  & YES & NA  & NA \\
{\bf OURS}                               & {\bf Green SE + Green AI} & {\bf 2010--2024}  & {\bf 390} & {\bf SMS}      & {\bf YES} & {\bf YES} & {\bf YES} & {\bf YES} \\
\bottomrule
\end{tabular}
\caption{Comparison of existing surveys and their coverage of the research questions addressed in this study. SMS stands for systematic mapping survey, DLC for development lifecycle, and SLR for systematic literature review. NA means not applicable or not available depending the context.}
\label{tab:green_se_green_ai_surveys}
\end{table}

{\bf Evolution and publication trends (RQ1).}
In line with our findings, previous surveys have also reported that
Green SE publications appear predominantly in
conference venues~\cite{marimuthu2017software,lee2024survey,verdecchia2023systematic}.
For Green AI, in particular, we find that workshops contribute fewer
publications than conferences and journals, which is consistent with
the observations of Verdecchia et al.~\cite{verdecchia2023systematic}.
Among the most prominent venues for recent studies are ICSE and
GREENS~\cite{marimuthu2017software,mourao2018green},
a finding that also aligns with our results.
In addition, we identify JSS as one of the main journal venues contributing to Green SE.

However, the reader should consider that publication trends may vary
depending on the inclusion and
exclusion criteria adopted by each survey,
as well as the type of venues considered,
and the amount of papers analysed.
For this reason, we have also find differences between our survey
and others.
For instance, Lee et al.~\cite{lee2024survey} reported a decrease in the
number of publications in 2014, while Mour\~ao et
al.~\cite{mourao2018green} observed a decline in 2016.
By contrast, we find that 2016 was the most prolific year in
terms of publications,
followed by a temporary decline.
Moreover, Penzenstadler et al.~\cite{penzenstadler2014systematic} reported 19
studies for 2011, whereas we find fewer.

In most recent years, starting from 2023, we observe a renewed up-tick in publications.
This rise is supported by the increasing number of AI publications, which \citet{verdecchia2023systematic} have observed from 2020 and onwards. 
In particular, 76 of their collected publications stem from 2020 onwards.

{\bf Application domains (RQ2).}
While the defined application categories vary based on the survey,
we also find some overlap. 
For instance, the top three categories according to \citet{penzenstadler2014systematic} are:
SE \& Lifecycle; Services, Mobile \& Cloud; Metropolitan Areas \& Housing.
\citet{mourao2018green} had Mobile and Cloud as the largest application domains.
The largest application domain for Green AI publications is Edge Computing and IoT.
Similarly, our results revealed Mobile as the largest application category, followed by AI. 
Unlike Green SE, Cloud and Mobile applications play a smaller role for  \citet{verdecchia2023systematic}.
Additionally, \citet{penzenstadler2014systematic} observed a large number of exploratory and solution papers, which could be comparable to the large number of optimization papers we found.
Finally, Rolnick et al.~\cite{rolnick2022tackling}
identified broad climate change solution domains,
including electricity systems and transportation, and so on.
However, here, we are interesting in application domains
of software systems, e.g., mobile, cloud, and AI.

{\bf Development lifecycle (RQ3).}
Most related surveys categorise existing studies
according to the software development lifecycle phases defined by SWEBOK
(e.g., requirements, architecture/design, development, and testing)~\cite{mourao2018green,penzenstadler2014systematic,marimuthu2017software,wolfram2017sustainability,lee2024survey}.
Following the same approach,
we also analysed the literature across these phases.
Our findings reveal that the development (or construction) phase has emerged as the most extensively studied area in recent years,
whereas architecture and requirements remain
comparatively underexplored in the analysed corpus.

However, these findings differ from several earlier surveys~\cite{mourao2018green,
penzenstadler2014systematic,
marimuthu2017software,
wolfram2017sustainability},
published before 2020.
For instance, Mour\~ao et al.~\cite{mourao2018green} reported that a great deal of studies focused on requirements (24\%) and design (21\%), while fewer studies refer to coding, i.e., construction (11\%).
Similarly, Penzenstadler et al.~\cite{penzenstadler2014systematic} found that most publications were concentrated in the areas of design, management, models and methods, process, quality, and requirements.
Along similar lines, Marimuthu et al.~\cite{marimuthu2017software} identified software requirements as the most frequently investigated topic (18 studies), followed by testing (17 studies) and maintenance (12 studies), whereas software construction (4 studies) and design (2 studies) seemed to be less investigated.
Likewise, Wolfram et al.~\cite{wolfram2017sustainability} highlighted software process, requirements, and design as the most researched areas of the software lifecycle.

One possible explanation for these different result is that
our dataset includes more recent publications (after 2020),
where research predominantly focuses on code optimisation and implementation-level techniques, i.e.,
relevant to the construction phase of the software lifecycle.
In this respect, our findings align with the observations of the most recent closely relevant survey (published in 2024) to ours, the survey of Lee et al.~\cite{lee2024survey}, who also found that the majority of recent studies belong to the construction phase.

Overall, this trend suggests a shift in research focus over time.
Earlier studies (published before 2020) primarily emphasised establishing sustainability as a software requirement, whereas more recent studies (published after 2020)
increasingly assume sustainability as an explicit requirement,
focusing on developing methods, techniques, and tools
for constructing sustainable software systems.

{\bf Empirical practices (RQ4).}
To our knowledge,
only the survey by \citet{lee2024survey}
summarises empirical practices in the field.
To group the surveyed articles, they proposed a fine-grained task categorisation across the software cycle,
including planning, requirements, design, implementation, testing, development, and maintenance.
By contrast, we group the articles in five broad categories key elements (hardware, measurement, stability, replication, data sets)
for conducting experiments on Green SE.
Finally, Xu et al.~\cite{xu2021survey}
present best practices for conducting experiments
with deep learning algorithms consuming less energy,

{\bf Overall.} Although parallels can be drawn between our findings and those of existing surveys, prior studies have also explored aspects that fall outside the scope of our analysis.
Therefore, the reader may consider the following surveys for more details on challenges~\cite{lee2024survey}, tools and frameworks~\cite{marimuthu2017software,lee2024survey}, and bibliometric analyses~\cite{penzenstadler2014systematic,mourao2018green,verdecchia2023systematic,marimuthu2017software}.

\section{Conclusion}
\label{section:conclusion}

In this paper, we carried out a systematic mapping study on research covering sustainability aspects stemming from software engineering venues. 
In total, we collected and analysed \papersRelevant publications.
Our analysis reveals that GREENS and JSS account for the highest number of Green SE publications among the 50 distinct venues identified (RQ1).
Among the publications examined, we found that the largest application domain that is addressed is \textit{mobile} (i.e., 152 publications).
For 2023 and 2024 there is an uptick in the popularity of AI applications considering sustainability, such as Deep Neural networks or Large Language Models (RQ2).
To address energy concerns, the most popular study types either propose optimization approaches or carried benchmarking studies to recommend energy-friendly implementations (RQ3), covering 53\% of the \papersRelevant publications.
Furthermore, we selected a subset of \papersExtract publications to extract details on empirical practices for measuring and evaluating energy in SE works (RQ4).
Among others, we observed a stratified field of measurement tools, with 31 unique ones being used by at least one of the \papersExtract publications.
Additionally, it is critical to consider the stability of experiments
and the potential influence of noise. 
To combat these, multiple runs have been conducted (e.g., 10 or 30) and warmup and cooldown steps are being taken.
Finally, RQ4 reveals that the field of Green SE requires researchers
to systematically report the settings of their studies to improve
replicability, as well as the introduction of new dedicated real-world datasets
for evaluating the sustainability of modern software systems.

\section*{Data availability}
To allow for reproducibility of our results, we make publicly available the labelled search results (RQ1-3) and the information extracted (RQ4).\footref{fn:shared}


\bibliographystyle{ACM-Reference-Format}
\bibliography{library,extraction}

\end{document}